# Introgression Browser: High throughput whole-genome SNP visualization

Saulo Alves Aflitos[1], Gabino Sanchez-Perez[1,2], Dick de Ridder[2], Paul Fransz[3], Eric Schranz[4], Hans de Jong[5], Sander Peters[1,*]

1. Applied Bioinformatics, Wageningen University and Research Centre (WUR), Wageningen, the Netherlands.

2. Bioinformatics Group, Wageningen University and Research Centre (WUR), Wageningen, The Netherlands.

3. Nuclear Organisation, Swammerdam Institute for Life Sciences, University of Amsterdam UvA). The Netherlands.

4. Biosystematics Group, Wageningen University, Droevendaalsesteeg 1, 6708 PB, Wageningen, The Netherlands

5. Laboratory of Genetics, Wageningen University and Research Centre (WUR), Wageningen, The Netherlands.

**Running title:** Introgression Browser

***Corresponding author**: Sander A. Peters; tel: +31-317-481123; fax: +31-317-418094; e-mail: sander.peters@wur.nl

**Word count**: Abstract 249 words; Introduction 839 words; Experimental procedures 1354 words; Results 928 words; Discussion 589 words; Acknowledgements 25 words; Short legends 78 words; References 1714 words; Figure legends 351 words; Total 6127 words.

# Abstract

Breeding by introgressive hybridization is a pivotal strategy to broaden the genetic basis of crops. Usually, the desired traits are monitored in consecutive crossing generations by marker-assisted selection, but their analyses fail in chromosome regions where crossover recombinants are rare or not viable. Here, we present the Introgression Browser (IBROWSER), a novel bioinformatics tool aimed at visualizing introgressions at nucleotide or SNP accuracy. The software selects homozygous SNPs from Variant Call Format (VCF) information and filters out heterozygous SNPs, Multi-Nucleotide Polymorphisms (MNPs) and insertion-deletions (InDels). For data analysis IBROWSER makes use of sliding windows, but if needed it can generate any desired fragmentation pattern through General Feature Format (GFF) information. In an example of tomato (*Solanum lycopersicum*) accessions we visualize SNP patterns and elucidate both position and boundaries of the introgressions. We also show that our tool is capable of identifying alien DNA in a panel of the closely related *S. pimpinellifolium* by examining phylogenetic relationships of the introgressed segments in tomato. In a third example, we demonstrate the power of the IBROWSER in a panel of 600 *Arabidopsis* accessions, detecting the boundaries of a SNP-free region around a polymorphic 1.17 Mbp inverted segment on the short arm of chromosome 4. The architecture and functionality of IBROWSER makes the software appropriate for a broad set of analyses including SNP mining, genome structure analysis, and pedigree analysis. Its functionality, together with the capability to process large data sets and efficient visualization of sequence variation, makes IBROWSER a valuable breeding tool.

# Introduction

Plant breeders apply introgressive hybridization for incorporating valuable traits from related species via interspecific hybridization and repeated backcrossings (Anderson, 1953). During meiotic prophase of these hybrids, crossover recombination may occur between the donor species and its homeologous counterpart of the recipient crop, thus creating novel allele combinations. Recombination usually involves the exchange of large fragments (up to complete chromosome arms), and can therefore result in the introgression of additional unwanted genes besides the target gene (Zamir 2001; Qi *et al.,* 2007). Breeders aim at maintaining the chromosomal regions of interest while removing the unwanted traits from the alien relative by recurrent backcrossing with the crop, followed by trait selection. The detection of introgressed segments harboring genes associated with economically important traits, whether or not linked to potentially deleterious or unwanted genes / alleles from the donor species, is therefore of eminent interest to breeders. Tracing of such introgressions in offspring families may also provide insight into homeologous recombination leading to incorporation of the desired genes from the alien donor into the recipient crop, while eventually losing genes controlling unwanted traits.

Detection of introgressed regions has previously relied extensively on diverse molecular markers technologies including Simple Sequence Repeats (SSR), Restriction Fragment Length Polymorphisms (RFLP), Amplified Fragment Length Polymorphisms (AFLP) and DNA microarrays (Rieseberg and Ellstrand*,* 1993; Powell *et al.,* 1996; Dekkers and Hospital, 2002; Víquez-Zamora *et al.*, 2013). However, the power of marker-assisted technologies to detect and delineate introgressions and chromosome rearrangements is limited due to low marker placement accuracy and even lack of markers (Kumar *et al.*, 2012; Víquez-Zamora *et al.*, 2013; Anderson *et al.*, 2011). In addition, sequence duplications, heterozygosity, and discrepancies between genetic and physical maps can seriously hamper data interpretation. Such limitations can partly be overcome through the use of *in situ* hybridization techniques such as Genomic *in situ* Hybridization (GISH) and Fluorescent in situ Hybridization (FISH). GISH technology can be used to obtain information on the size and number of alien chromosomes or chromosome segments, interspecific and intergeneric translocations resulting from homeologous recombination, and the presence and approximate location of introgressed genes (Schwarzacher *et al.*, 1992; Thomas *et al.*, 1994; Chang and de Jong, 2005). In a combined chromosome painting with GISH followed by FISH using BAC DNA as probes, Dong and coworkers (2001) even revealed genetic identity of alien chromosomes and segments in potato breeding lines. Such strategies, however, are not sufficient for unravelling complex rearrangements and identification of chromosome breakpoints at nucleotide accuracy. To this end we introduced the combined usage of BAC FISH, genetic markers and reference genome sequence information to elucidate the complex topology of tomato and potato chromosomes as well as detection of introgressed regions (Tang *et al.*, 2008; Peters *et al.*, 2009; Peters et al., 2012; Aflitos *et al.*, 2014).

Other methods for introgression detection include Restriction site Associated DNA (RAD) and Genotyping by Sequencing (GBS), which both generate restriction fragments that can be subsequently sequenced for posterior SNP calling (Baird *et al.*, 2008). GBS allows high-throughput detection of thousands of SNPs along the genome, producing a polymorphism density several orders of magnitude higher than genetic marker based technologies, but is also two orders of magnitude less sensitive than next-generation sequencing (NGS). Furthermore, the resolution depends on the occurrence of restriction sites, which are not evenly distributed along chromosomes. Due to the dependency on enzymatic activity, there is a high rate of non-calls and relatively low reproducibility that makes GBS less suitable for introgression detection (Galvão *et al.*, 2012).

The examination of Genome-wide Single Nucleotide Polymorphisms (SNPs) is an alternative strategy that becomes increasingly attractive for disclosing genome organization and topological context of target genes underlying agronomically important traits. Several methods to visualize such whole-genome SNP (wgSNP) data have been presented (Posada, 2002; Martin *et al.*, 2011; Lechat *et al.*, 2013; Kim *et al.*, 2014) including software packages such as VISRD (Strimmer *et al.*, 2003), SHOREMAP (Schneeberger *et al.*, 2009), RECOMBINE (Anderson *et al.*, 2011), NGM (Austin *et al.*, 2011), PARTFINDER (Prasad *et al.*, 2013) and PHYLONET-HMM (Liu *et* *al.*, 2014). SHOREMAP and NGM focus on identifying causal SNPs for phenotypes in segregating populations. VISRD, PARTFINDER and PHYLONET-HMM are based on a sliding window approach by mapping different genomes to a reference genome and searching for inconsistent phylogenetic relationships that are used as markers for introgressions. However, visRD and ReCombine only handle relatively small genomes, whereas SHOREmap, NGL and PHYLONET-HMM are targeting inbred lines. PARTFINDER analyses several large genomes and creates global phylogenetic trees, however, it does not identify introgressed segments. In this paper we present the Introgression Browser (IBROWSER), a tool that delineates introgressed segments, identifies donor parents, and is able to handle a large number of genomes with virtually no genome size limitation. We have separated the computationally intensive database calculations from the visualization and use open standards, allowing the reuse of the data and data exchange with other supporting programs. IBROWSER fills the technological gap between high-throughput sequencing and sequence-based introgression detection, and is applicable for large or industrial scale introgression hybridization breeding programs.

## Experimental procedures

IBROWSER consists of a back-end which calculates and stores SNPs in a database and a front-end that enables the visualization of SNPs. IBROWSER takes SNP data from any variant calling algorithm provided in VCF format, and can import distance matrix information from multiple programs such as FASTTREE (Price *et al.*, 2010) or SNPPHYLO (Lee *et al.*, 2014). This provides the user flexibility to operate and test combinations of parameters for proper visualization. IBROWSER is able to efficiently process and filter data with excessive noise, although we do advise the use of high quality SNP calls and repeat masking. The set of scripts and programs necessary to run the IBROWSER can either be downloaded as a pre-configured virtualized disk bundle or can be installed directly in a pre-existing system.

## Data Generation

To visualize introgressions in wild *Solanum* species and crop tomatoes, we created three datasets "84-10k", "84-50k" and "84-Genes" from the genome sequence information produced for 84 tomato accessions (Aflitos *et al.*, 2014). The 84-10k and 84-50k datasets contain consecutive segments of 10 kbp and 50 Kbp, respectively. The 84-Genes dataset consists of segments generated from the *S. lycopersicum* cv. Heinz v2.40 using gene coordinates from the v2.31 annotation (Tomato Genome Consortium, 2012). Approximately 5 hours using a 20 core Intel(R) Xeon(R) CPU E7- 4840 @ 2.00GHz was required to generate each set. The visualization takes up to 20 seconds for the largest dataset.

A dataset, hereafter referred as "60-IL" dataset, was created consisting of tomato, *Solanum lycopersicum* cv. MoneyMaker LYC 1365, *S. pimpinellifollium* CGN14498, 60 offspring F6 Inbred Line (IL) individuals, and four *S. pimpinellifolium* related accessions (LYC2798, LYC2740, LA1584 and LA1578). The tomato Moneymaker parent, 60 ILs and the *pimpinellifolium* related accessions were obtained from Aflitos *et al.*, (2014). The *S. pimpinellifollium* CGN14498 parent was sequenced using a 500 bp paired-end library with the Illumina HiSeq 2000 platform at 38 fold coverage (assuming a genome size of 950 Mbp). The 60 F6 IL individuals were likewise sequenced at 4.80±1.70 fold coverage. *S. pimpinellifollium* CGN14498 and the 60 IL individuals were mapped against *S. lycopersicum* cv. Heinz 1706 v2.4 as described in (Aflitos *et al.*, 2014) using BWA (Li *et al.*, 2009a) and SNPs were called using SAMTOOLS (Li *et al.*, 2009b). The raw data of the 60 RILs and *S. pimpinellifollium* CGN14498 were deposited at The European Bioinformatics Institute (http://www.ebi.ac.uk/) under the PRJEB6659 identification number. The resulting VCF file was split into non-overlapping 50 Kbp fragments.

**The back-end**

The back-end consists of a database creation tool that takes Variant Call Format (VCF) files as input generated from samples that are mapped to a reference genome, and a FASTA file containing the reference genome sequence. It converts VCF and FASTA data into an intermediate file format containing the genotype for each coordinate per sample. Alternatively, this can also be achieved by using SNP-SEARCH (Al-Shahib and Underwood 2013). Coordinates are excluded if any of the following constraints are not satisfied; (i) at least two individuals show polymorphism; (ii) any individual shows heterozygosity; or (iii) any individual contains an InDel, a protocol modified from Lee and coworkers (2014). We filter out InDels because they are difficult to align and score (Anderson *et al.*, 2011). The reason is that they are generally not fixed in the population and are usually of little value for marker-assisted breeding. Furthermore, they cannot be expressed in canonical FASTA format and most phylogeny programs do not accept IUPAC codes for ambiguous nucleotides. The filtering of polymorphisms usually reduces the input data size up to 30-40%.

The remaining coordinates are then ordered and connected head to tail into consecutive fragments. Alternatively, IBROWSER can operate in a sliding window mode where each fragment overlaps with the previous. The latter operating mode allows for more accurate introgression boundary detection. However, it will generate significantly more data compared to the consecutive operating mode, and the overlapping mode therefore should be used exclusively for a limited number of small segments.

Any General Feature Format (GFF) file containing segmentation information can be used, such as coordinates of CDSs, genes, exons or QTLs. If no particular segmentation pattern is specified, the reference genome FASTA file may be used to create a GFF file containing a user-defined, evenly spaced, segmentation pattern (Huang *et al.*, 2009; Anderson *et al.*, 2011; Prasad *et al.*, 2013; Chen *et al.*, 2014; Kim *et al.*, 2014, Liu *et al.*, 2014). The resulting GFF is then used to split the VCF files into the desired segments, as consecutive windows, sliding windows, or any arbitrary filtering from the annotation file of the reference genome such as CDS, genes, exons, introns, etc. Finally, for each segment a column is produced containing the polymorphism count for each sample.

All SNPs for each segment are then concatenated per species and stored as FASTA files. Next, a Jukes-Cantor distance matrix and a Newick formatted Maximum Likelihood phylogenetic Bio-Neighbor Joining tree are constructed per coordinate using FASTTREE2 (Price *et al.*, 2010). The distance matrices and Newick formatted phylogenetic trees are then parsed and stored into a portable database. Alternatively, the process of filtering, exporting and processing the VCF files can be performed using SNPPHYLO (Lee *et al.*, 2014).

In the case of Introgression Lines (ILs) we assume the genotype of an offspring individual originates either from the donor or acceptor parent and, following this paradigm, we therefore assign each SNP to either parent. To achieve this, we first calculate the average number of SNPs per sample, subsequently test whether the number of SNPs is above or below the average and then convert the sequence of the sample to the sequence of the closest parent according to the method of Kim *et al.*, (2014) and Huang *et al.*, (2009). In this way miscalls due to low genomic coverage are repaired, resulting in improved accuracy for delineated introgressed segments and a more reliable donor species identification.

Finally, we store the phylogenetic files (Newick tree and distance matrix) either as a PYTHON memory dump, which allows the information to be stored in RAM memory for fast access and high request loads, or as an SQLITE database. The latter takes a few milliseconds longer to serve the results, but has a smaller memory footprint. The database file independently can be copied to any other computer running the front end.

**The front-end**

Using the project database, the front-end generates the UI in the web browser. The required Python web server can run in any OS, which is able to run Python. It can be accessed locally through the intranet, or via the internet, depending on the configuration of the host computer. As the front-end employs HTML5 technology to display the User Interface (UI), no plugin installation is required. An optional user access control system is available providing security at the login level.

For each fragment the IBROWSER UI displays the pairwise distance values between the reference and each query sample as colors in a heat map, composed of lines and columns representing samples and genomic segments respectively. Any sample from a panel of genomes can be selected as the query sample. The first line of the heat map for each segment shows the number of SNPs using a contrasting color scale, giving insight into the distance (Figure 1).

The color scale for the pairwise distances as well as the color scale for the number of SNPs is displayed at the top of the graph. The heat map color scheme and row identifiers (samples) can be customized for better discrimination between samples. Clicking in the heat map shows a tooltip box for the selected cell including its statistics on SNP counts and distances. Double clicking in the heat map or selecting a gene/fragment by name in the main menu invokes an overlay panel showing the complete data for the selected column. This includes the respective fragment identifier, coordinate, phylogenetic tree, nucleotide sequence and distance matrix, all of which can be downloaded in their native formats. The full heat map can be downloaded as an PNG/JPEG image. The web server has a JSON API allowing developers to retrieve the data in an interchangeable format, facilitating cross-program usage.

# Results

**iBrowser shows homeologous recombination sites in introgression lines**

To analyse the chromosome structure of the 60 introgression lines (ILs) we aligned the sequence reads to both the *S. lycopersicum* Moneymaker and the *S. pimpinellifolium* CGN14498 parent and display the sequence distance to either parent in a heat map. Figure 2 displays the Tomato chromosome 6 heat map for all ILs. Crossover sites denote positions where chromosome origin changes from one parent to the other. This is displayed by a color change in the heat map. Clearly, about half of the individuals hardly differ from the MoneyMaker parent across most of the chromosome (indicated by a red box in the figure), and hence have a high sequence similarity almost without any introgression from *S. pimpinellifolium*. One other quarter shows a high distance (green box) to MoneyMaker; the SNP pattern suggests that almost the entire Chromosome 6 was replaced by its homeologous part of *S. pimpinellifolium*. The heat map using the *S. pimpinellifolium* parent as a reference (Figure 2, panel B) confirms that regions with a high distance to MoneyMaker are identical to the *S. pimpinellifolium* parent, generating an almost inverted image compared to the heat map in Figure 2, panel A. The sequence homology between the parent *S. pimpinellifolium* CGN14498 and its *S. pimpinellifolium* relatives (LYC2798, LYC2740, LA1584, LA1578, and CGN14498) is very high (Figure 2, panel A, yellow box), with CGN14498 being the closest match. This result illustrates the ability to identify the source of an introgression from a panel of very closely related *S. pimpinellifolium* individuals.

The remaining samples show intermediate distances to MoneyMaker (outside the boxes). Most likely this is caused by the extremely low coverage of sequencing (approximately 6x), yielding insufficient evidence for SNP calling. Normally, whenever lack of coverage occurs, the reference sequence is assumed. Because such cases hamper identification of the source of introgression (Canady *et al.*, 2006; Anderson *et al.*, 2011), we included a filter specifically designed to handle ILs with low coverage sequencing. It assumes that the genotype of the offspring is identical to either the recipient or donor parent (see methods). Figure 2, panel C shows the result for the 60-IL dataset after such a IL-specific pre-treatment of the data where each segment was assigned to either parent, without taking intermediate values. To delineate an introgression at base pair level accuracy, the SNP alignment for the segment containing the crossover has to be analyzed in detail in order to identify the crossover site.

**iBrowser reveals positions of alien introgressions**

Previously we manually traced introgressions in tomato (*Solanum lycopersicum*) crop accessions from related wild species based (Aflitos *et al.*, 2014). Here we used the iBrowser pipeline to test the automated detection and visualization of introgressions. Figures 3 and S1 visualize the introgressions from *S. pimpinellifolium* LYC2798 in *S. lycopersicum* cv. MoneyMaker LA2706 and *S. lycopersicum* cv. AllRound LA2463 in chromosome 6. The phylogenetic tree for the segment reveals the clustering of both Moneymaker and AllRound with the *S. pimpinellifolium* clade, which is the likely source of the introgression. Moneymaker and Allround group with the other *S. lycopersicum* accessions for positions that flank the right and left borders of the segment delineating the size of the introgression (Supplementary Figure S2). This finding is consistent with our previous results (Aflitos *et al.,* 2014).

The average gene size in tomato is approximately 3.7kb (Aoki *et al.*, 2010), and the number of SNPs in genes between tomato cultivars is generally smaller than that in 10 kbp or 50 kbp windows. The lower SNP content from genic sequences obtained by exome capture nevertheless has been shown to contain sufficient information for solid phylogenetic analysis (Austin *et al.*, 2011; Galvão *et al.*, 2012). In addition, our previous findings indicate that SNPs in coding sequences of *Solanum* accessions are under higher selective pressure than SNPs in non-coding regions and contain more phylogenetic information than non-coding SNPs (Aflitos *et al.*, 2014). Supplementary Figure S3 shows the analysis on the 84-10k, 84-50k and 84-Genes datasets. The latter dataset, in which each annotated tomato gene is an individual fragment, visualizes regions of higher conservation across the whole tomato clade as well as introgressions that are shared exclusively by hybrid tomatoes such as beefy and cherry varieties. These results show that IBROWSER is able to detect accession specific introgressions that can be used for pedigree analysis studies.

**iBrowser for revealing aberrant SNP landscapes in and around inversions**

The short arm of chromosome 4 in the *Col*-0 and few other accessions contains a 1.17 Mb paracentric inversion (Fransz *et al.*, manuscript in prep). In and around this chromosomal rearrangement there is almost complete absence of SNPs in contrast to the syntenic segment in Landsberg *erecta* (*Ler*) and most other accessions. Also in the pericentric heterochromatin the group of accessions containing the inversion do not have SNPs. This is shown by sequence analysis of 596 *Arabidopsis thaliana* accessions from the 1001 Arabidopsis database (Cao *et al.*, 2011; Schmitz *et al.*, 2013; Wang *et al.*, 2013a) and clearly illustrated by the heatmap of this segment in iBrowser (Figures 4 and S4). Accession hybrids heterozygous for the inversion will not produce viable gametes with recombinations in the inverted region and so display absence of recombinants in the genetic map (Drouaud *et al.*, 2006; Wang *et al.*, 2013b) and hence hardly any SNPs in the inversion and its flanking regions. Our SNP detection results are consistent with these previous observations. These results illustrate that the parallel visualization of introgression patterns along chromosomes in iBROWSER may help breeders to identify regions which are problematic for introgression hybridization breeding.

# Discussion

SNPs are becoming increasingly important in establishing source identity of introgressions (Huang *et al.*, 2009; Chen *et al.*, 2014; Paraskevis *et al.*, 2005; Anderson *et al.*, 2011; Austin *et al.*, 2011; Víquez-Zamora *et al.*, 2013; Prasad *et al.*, 2013). Plant genomes, also those of crop species, contain massive amounts of SNPs in their gene pools. Our study on genetic variation shows the number of SNPs in tomato (*S. lycopersicum*) and related wild species can exceed 10 million (Aflitos *et al.*, 2014). Previously, we used genome-wide SNP information and phylogenetic analyses to detect the source of introgressed segments in crop tomato. However, open source software tools that integrate SNP detection, visualization and phylogenetic analysis in a single workflow aiming at source identification of introgressed segments on a genome wide scale were lacking. Here, we have developed iBrowser, which overcomes this challenge. The software requires only a small memory and disk footprint, permitting it to run on any modern computer platform, while its databases can be distributed easily. It is also highly portable and can be used in stand-alone mode or over a network connection to facilitate project sharing. IBROWSER can be accessed through its JavaScript Object Notation (JSON) Application Programming Interface (API), allowing it to combine with programs such as MUMMERPLOT (Kurtz *et al.*, 2004) and (G/J)BROWSE (Stein *et al.*, 2002; Skinner *et al.*, 2009). This opens the possibility to integrate additional information on chromosome structure, which can help to assist in breeding parent selection. The software is freely distributed at https://github.com/sauloal/introgressionbrowser under the MIT license.

We implemented a new computational approach to process sequence information from a large panel of genomes and retrieve information on the location, size, and source of introgressed segments. We show that IBROWSER is capable of detecting characteristic introgressions in cherry and beef tomato accessions. In addition, by including the ITAG2.31 annotation data for tomato into IBROWSER, the genic content of introgressed segments can be retrieved, which may help to find additional leads on gene traits relationships. Furthermore, this bioinformatics tool identified the most likely donor of a 2.2 Mbp introgressed segment into the *S. lycopersicum* cv.

Moneymaker recipient from a panel of closely related *S. pimpinellifolium* accessions. This functionality makes IBROWSER an applicable tool for pedigree analysis.

We also used IBROWSER to analyze a large panel of Introgression Lines. By applying a specific filter we removed erroneous SNP calls that cannot be assigned to either breeding parent, and so improve the accuracy of SNP detection and the identification of introgressed segments even in genomes that have been sequenced at 4X coverage. This capability provides a new perspective for molecular breeding and allows studying genome features involved in crossover recombination in more detail. The information on size and sequence of introgressed segments, and frequency of crossover recombination that can be inferred from the SNP distribution, may be used to search for signatures that can either promote or prevent crossover recombination. Such features might also include information on genome structure either promoting or preventing crossover recombination. Using a comparative analysis on a panel of 600 accessions, we could pinpoint a 1.17 Mb inverted segment in the top arm of chromosome 4S that has been implicated in absence of crossover recombination in inversion heterozygotes (Drouaud *et al.*, 2006). In addition to this large inversion, IBROWSER revealed many other segments that were devoid of SNPs. It will be interesting to investigate the relationship between the SNP distribution and topology of these segments. These results illustrate the ability of IBROWSER as a tool to assist in comparative genomics studies.

# Acknowledgements

We thank Suzanne Hoogstrate, Luca Santuari and Eric Schranz for comments in the manuscript; this research was supported by the Centre for BioSystems Genomics (CBSG).

## Short legends for supporting information

**Figure S1:** Heat map for tomato chromosome 6 from 84 *Solanum* accessions.

**Figure S2:** Maximum Likelihood phylogenetic trees for 50 Kbp segments from a chromosome 6 introgression.

**Figure S3:** Heat maps plotted with *S. lycopersicum* cv. Heinz as reference and ordered as in figure S1, using the 84-10kb (panel A), 84-50kb (panel B) and 84-Genes (panel C) datasets respectively.

**Figure S4:** Heat map for a fragment of the top arm of chromosome 4 for 596 *Arabidopsis thaliana* accessions.

Koornneef, M., Keurentjes, J.J., Schneeberger, K. (2013) The genomic landscape of meiotic crossovers and gene conversions in *Arabidopsis thaliana*. *eLife*, e01426.

**Zamir, D.** (2001) Improving plant breeding with exotic genetic libraries. *Nature Rev. Gen.* **2**, 983-989.

## Figure legends

**Figure 1:** User interface of the IBROWSER. Panel A shows the menu for database, reference species, chromosome, gene and grouping selection; in panel B the first line shows the alignment statistics; the second line presents interactive buttons for download, toggling of row names, clustering method selection, zoom level and heat map color scheme selection; in panel C the first line displays the color scales used for the phylogenetic distance, the number of SNP per segment, and the information of the selected database and filters; and in panel D the heat map is shown in black and white, with the sample names on the left side, a ruler at the top and a yellow-red scale representing the number of SNPs in each segment.

**Figure 2:** Heat maps for 60-RIL samples (*y*-axis, same order in all panels) which are mapped against *S. lycopersicum* cv. Heinz chromosome 6 (*x*-axis) with a 50kpb window size. Panel A displays distances to *S. lycopersicum* cv. MoneyMaker; panel B displays same data as shown in panel A (though vertically compressed) using a RIL-specific data filter; panel C displays the distances to *S. pimpinellifolium* CGN14498. Red and green boxes highlight RILs with a large introgression from *S. lycopersicum* cv. MoneyMaker or S. *pimpinellifolium* respectively. Rows highlighted in yellow are the *S. pimpinellifolium* samples. Rows without highlighting are RIL individuals with a low SNP coverage, showing intermediate distances.

**Figure 3:** Heat maps for tomato chromosome 6 between coordinates 36 Mbp and 40 Mpb, using *S. lycopersicum* cv. MoneyMaker (panel A) and *S. pimpinellifolium* LYC2798 as reference (panel B) respectively. The segment originates from *S. pimpinellifolium* LYC2798 that is introgressed into *S. lycopersicum* cv. MoneyMaker and *S. lycopersicum* cv. AllRound as has previously been described in Aflitos *et al.*, (2014). Individuals with the introgression are highlighted in red.

**Figure 4:** Fragment of the top arm of chromosome 4 for 596 *Arabidopsis thaliana* accessions obtained from the 1001 Arabidopsis consortium. White blocks show reduction in the number of SNPs, coinciding with an inverted segment implicated in absence of recombination. The few SNPs that are still present may have occurred from mutation events.

**Figure 1:** User interface of the iBROWSER. Panel A shows the menu for database, reference species, chromosome, gene and grouping selection; in panel B the first line shows the alignment statistics; the second line presents interactive buttons for download, toggling of row names, clustering method selection, zoom level and heat map color scheme selection; in panel C the first line displays the color scales used for the phylogenetic distance, the number of SNP per segment, and the information of the selected database and filters; and in panel D the heat map is shown in black and white, with the sample names on the left side, a ruler at the top and a yellow-red scale representing the number of SNPs in each segment.

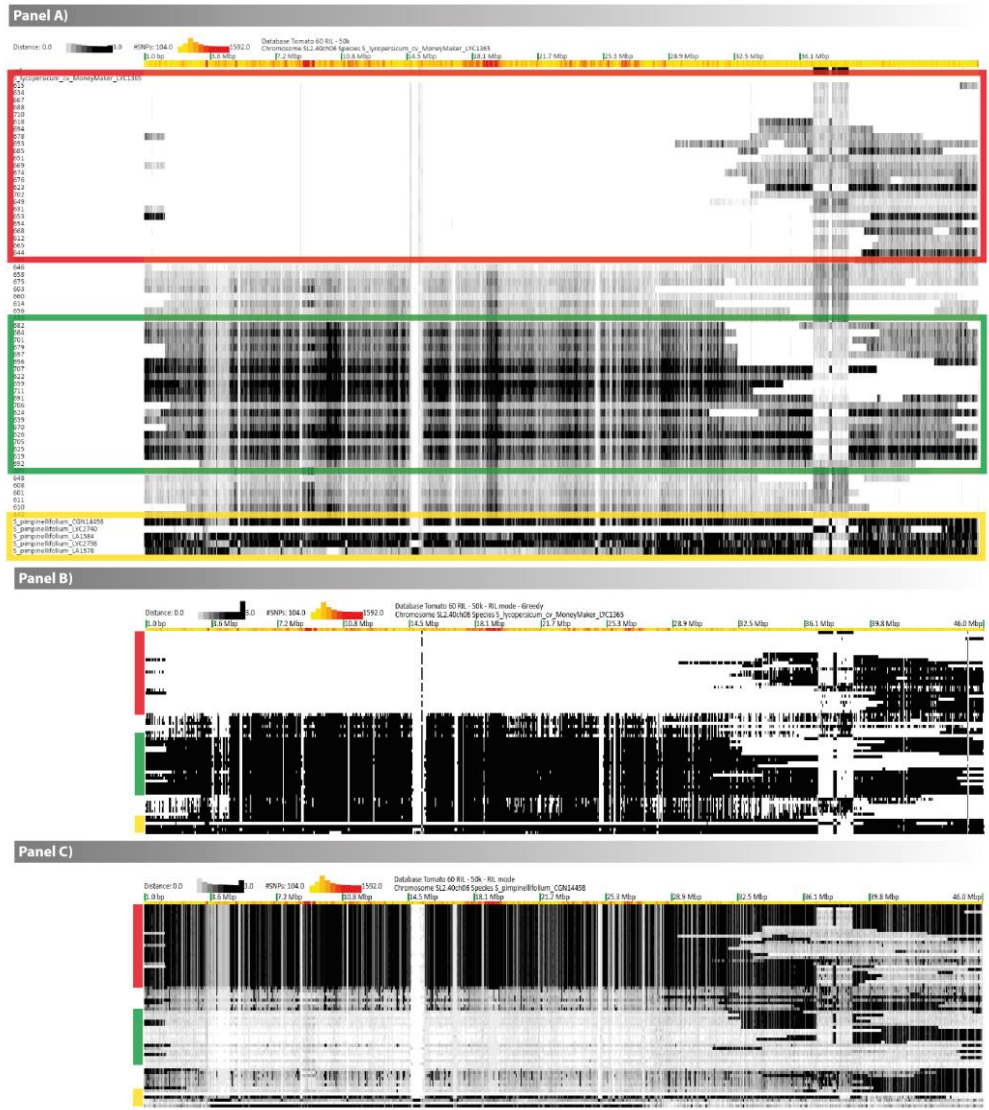

**Figure 2:** Heat maps for 60-RIL samples (*y*-axis, same order in all panels) which are mapped against *S. lycopersicum* cv. Heinz chromosome 6 (*x*-axis) with a 50kpb window size. Panel A displays distances to *S. lycopersicum* cv. MoneyMaker; panel B displays same data as shown in panel A (though vertically compressed) using a RIL-specific data filter; panel C displays the distances to *S. pimpinellifolium* CGN14498. Red and green boxes highlight RILs with a large introgression from *S. lycopersicum* cv. MoneyMaker or S. *pimpinellifolium* respectively. Rows highlighted in yellow are the *S. pimpinellifolium* samples. Rows without highlighting are RIL individuals with a low SNP coverage, showing intermediate distances.

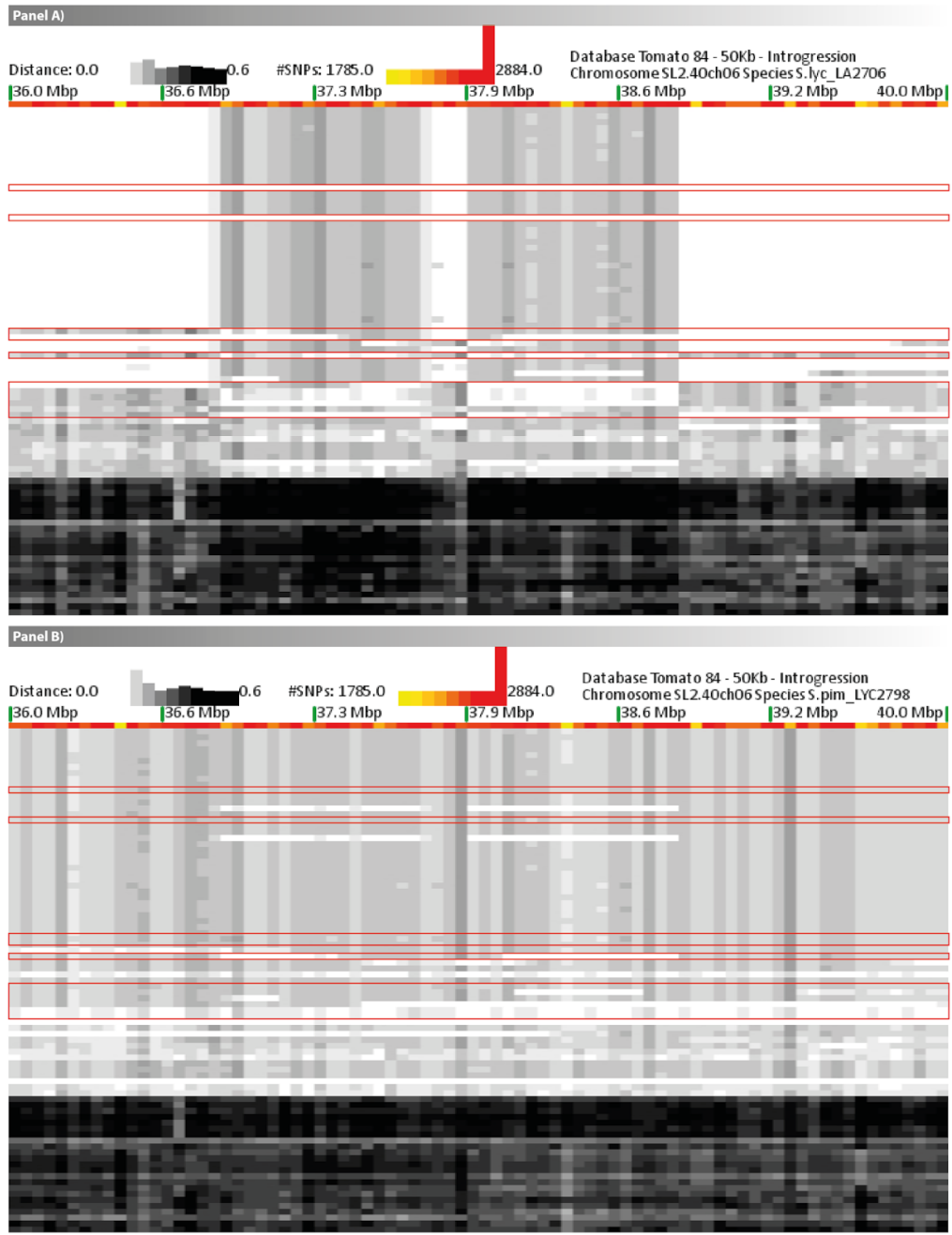

**Figure 3:** Heat maps for tomato chromosome 6 between coordinates 36 Mbp and 40 Mpb, using *S. lycopersicum* cv. MoneyMaker (panel A) and *S. pimpinellifolium* LYC2798 as reference (panel B) respectively. The segment originates from *S. pimpinellifolium* LYC2798 that is introgressed into *S. lycopersicum* cv. MoneyMaker and *S. lycopersicum* cv. AllRound as has previously been described in Aflitos et al., (2014). Individuals with the introgression are highlighted in red.

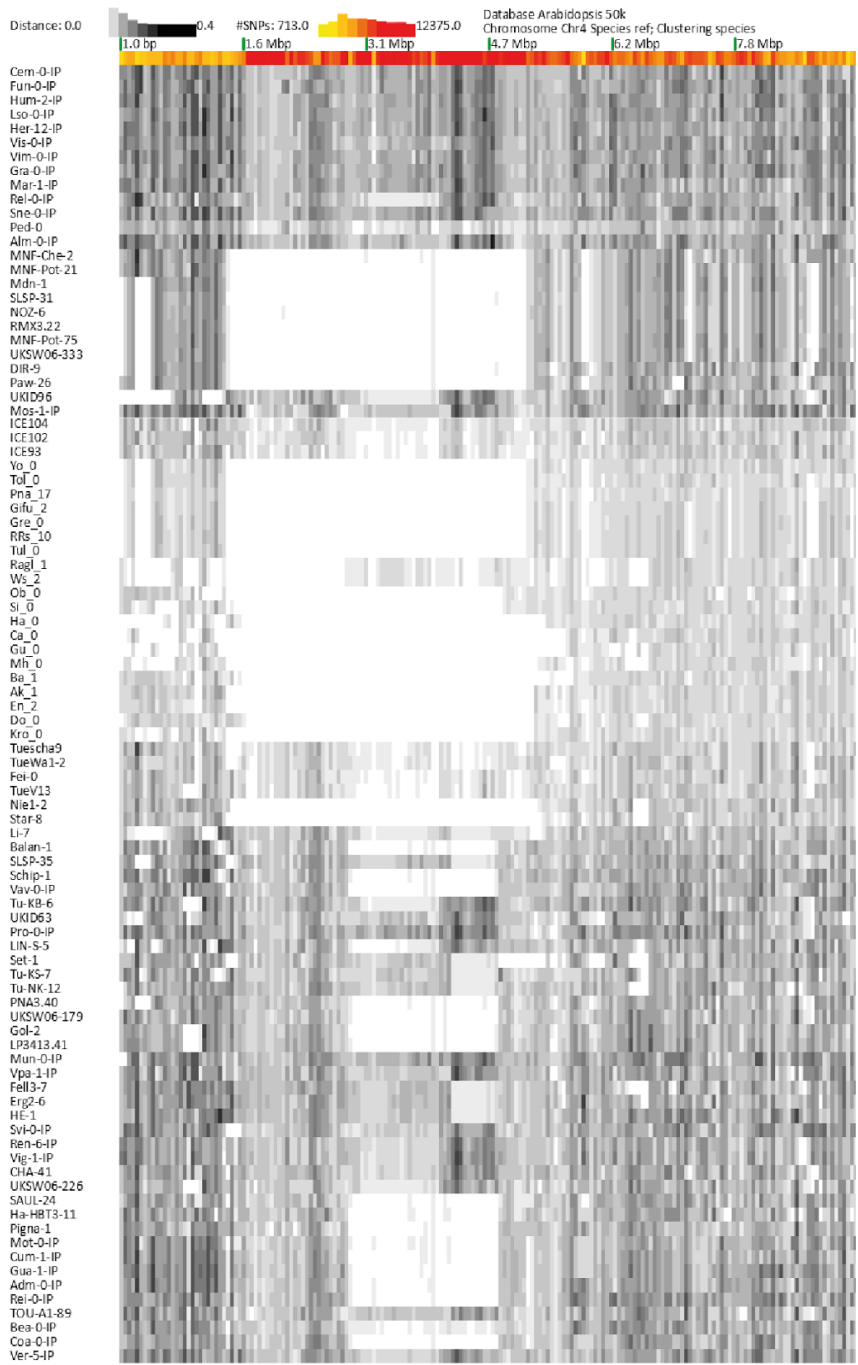

**Figure 4:** Fragment of the top arm of chromosome 4 for 596 *Arabidopsis thaliana* accessions obtained from the 1001 Arabidopsis consortium. White blocks show reduction in the number of SNPs, coinciding with an inverted segment implicated in absence of recombination. The few SNPs that are still present may have occurred from mutation events

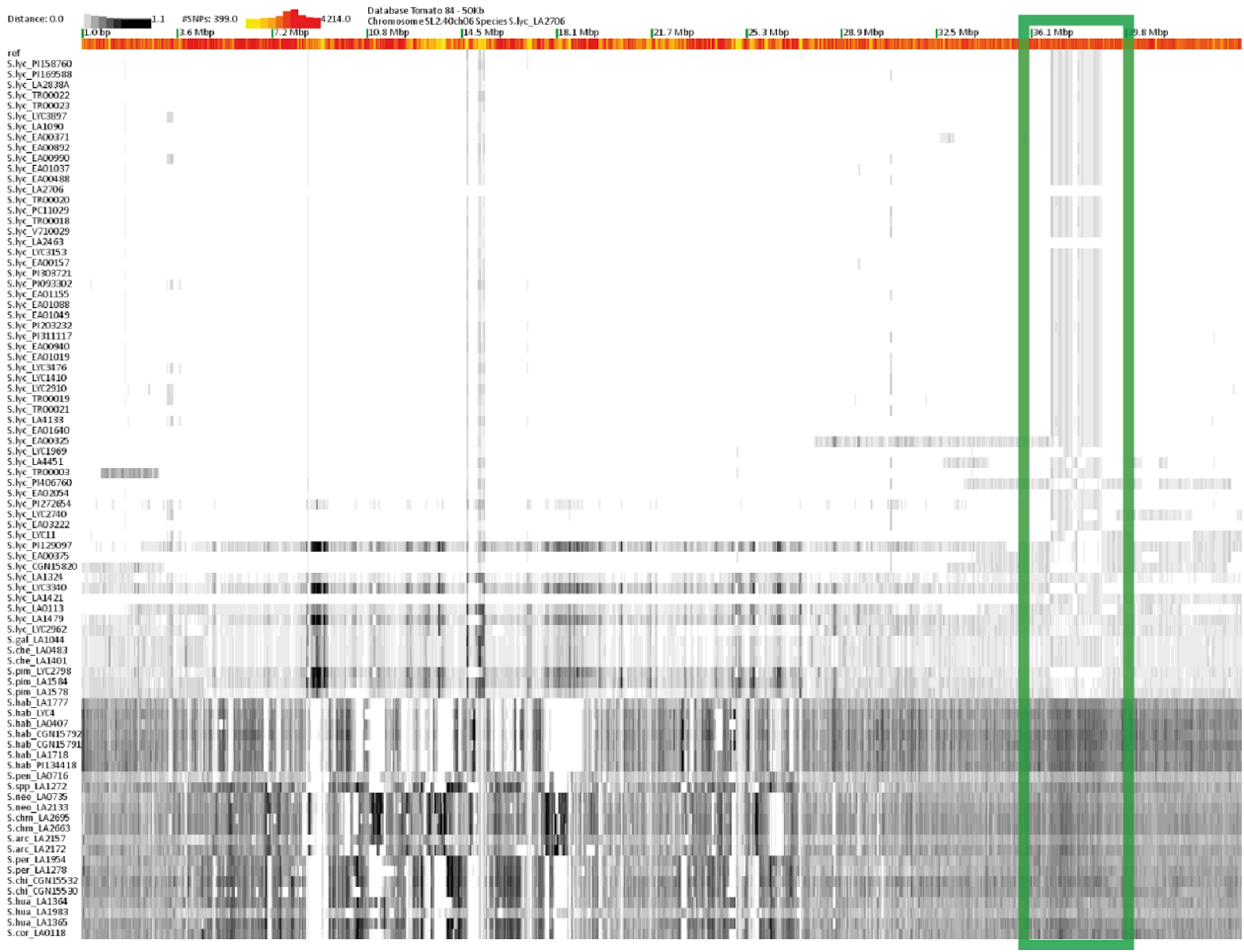

**Figure S1:** Heat map for tomato chromosome 6 from 84 *Solanum* accessions using *S. lycopersicum* cv. MoneyMaker as reference. Highlighted in green is the segment displayed in Figure 3. Each segment represents 50 Kbp in size.

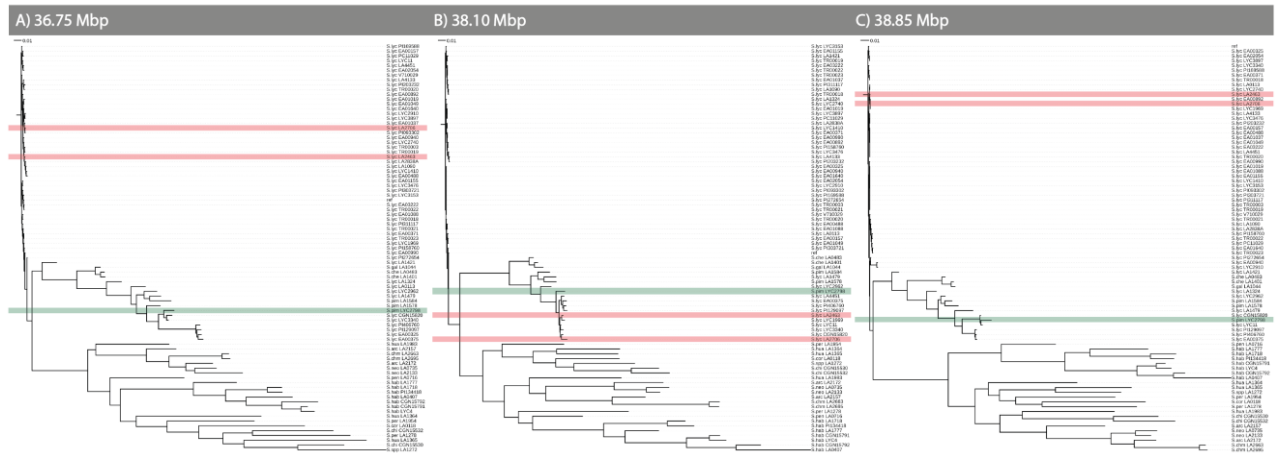

**Figure S2:** Maximum Likelihood phylogenetic trees for 50 Kbp segments from a chromosome 6 introgression positioned at the left side (panel A), inside (panel B), and at the right side (panel C) of the introgressed segment. The phylogenetic position of *S. lycopersicum* cv. MoneyMaker LA2706 and *S. lycopersicum* cv. AllRound LA2463 among other *S. lycopersicum* group accessions are highlighted in red and change their relative position to the closest relative donor species *S. pimpinellifolium* LYC2798 highlighted in green. Trees have been drawn using INTERACTIVE TREE OF LIFE v2.2.2 (Letunic and Bork, 2011).

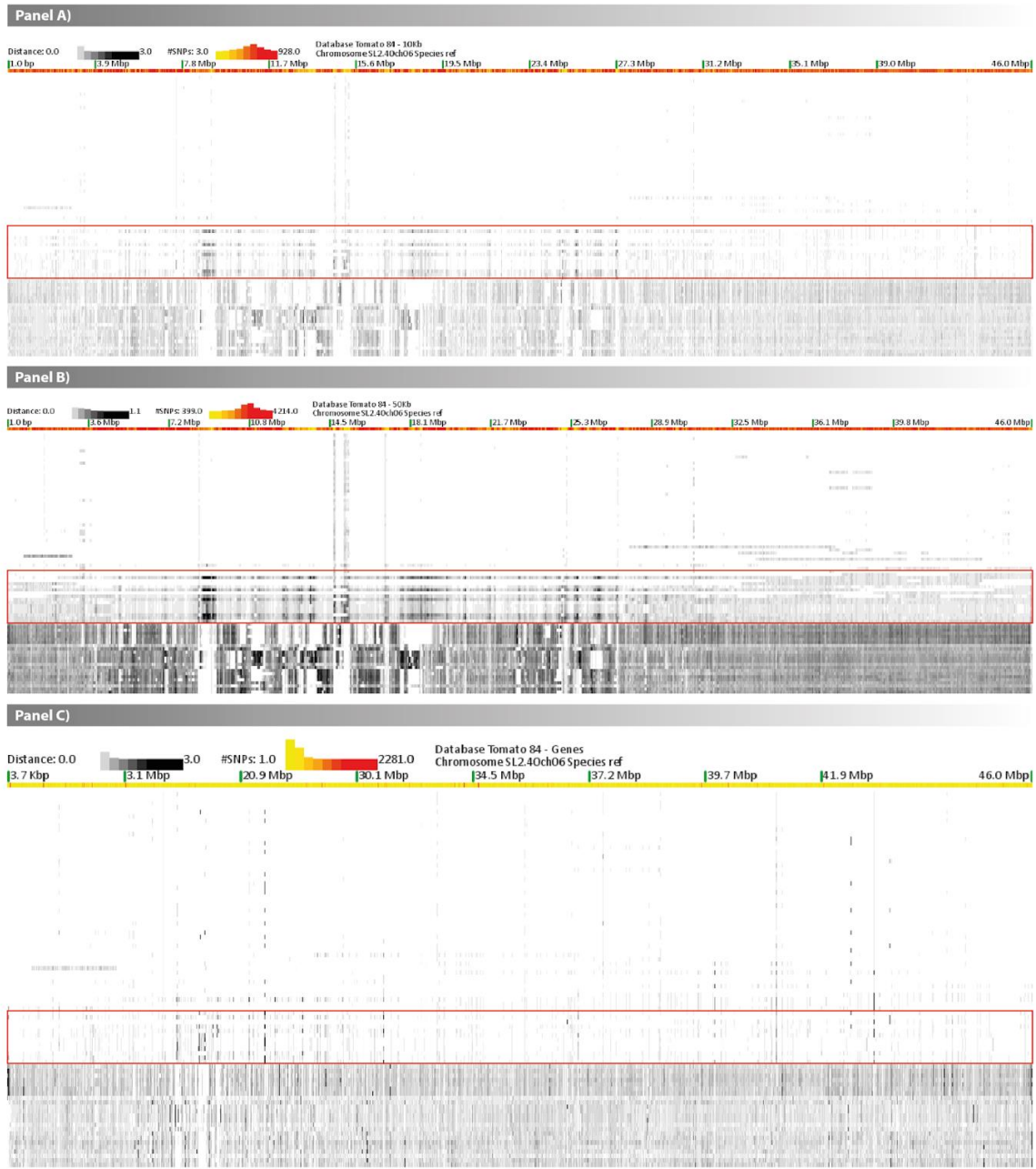

**Figure S3:** Heat maps plotted with *S. lycopersicum* cv. Heinz as reference and ordered as in figure S1, using the 84-10kb (panel A), 84-50kb (panel B) and 84-Genes (panel C) datasets respectively. Red blocks show clustering of wild *S. lycopersicum* and *beefy/cherry* tomato varieties.

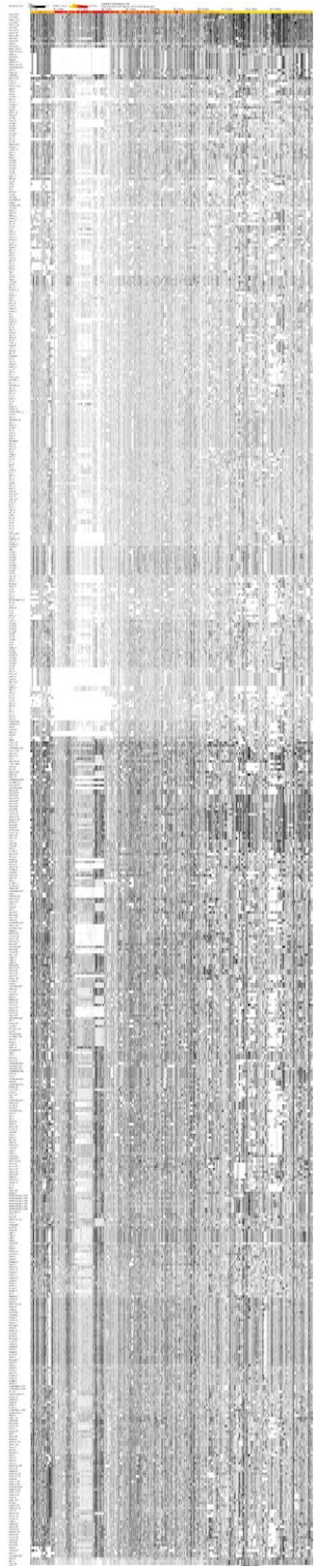

**Figure S4:** Chromosome 4 of 596 *Arabidopsis* accessions mapped against TAIR v 10 with 50 Kbp segments.